# Modeling and Performance Analysis of 10 Gbps Inter-Satellite-Link (ISL) In Inter-Satellite Optical-Wireless Communication (IsOWC) System between LEO and GEO Satellites

Tayyab Mehmoood
NUST, School of Electrical Engineering and Computer Science
Islamabad, Pakistan
tayyabjatoibaloch@gmail.com

Nauman Hameed
Dept. of Telecommunication
University of Engineering & Technology
Taxila, Pakistan
Nauman112te@gmail.com

*Abstract*— Free space optical communication has merged the aspects of fiber optics and the wireless communication which are the most conquered and controlled telecommunication technologies. Most of the features of free space optics (FSO) are interrelated to fiber optics but the difference between them is transmission medium, which is glass in case of fiber-optics and air/vacuum in case of FSO. In the near future, communication between LEO & GEO satellites with each other which are orbiting the Earth will be done by using inter-satellite optical wireless communication (IsOWC) systems. IsOWC systems is the most significant application of the FSO and it will be installed in the space in the near future because of its low input power, no licensing by ITU, low cost, light weight, small size of the telescopes and very high data rates as compared to the radio frequency (RF) satellite systems. In this research article, IsOWC system is designed between LEO and GEO satellites by using OPTI-System simulator which is not stated in past examined research works. Inter-satellite link is established between satellites which are separated by the distance of 40,000 Km at the bit rate of 10 Gbps.

*Index Terms*—Free space optics, optical wireless communication, inter-satellite link, inter-satellite optical wireless communication system.

## I. INTRODUCTION

Since the birth of human race, there has been a necessity of delivering messages from point A to remote point B with reliability, security and speed of transferring message is of extreme importance. Later on, it was acknowledged that the fastest way of communication was not a fastest horse but light. Hence, a network of light towers was created on the top of mountains in ancient Greece. Innovations and advancements in the field of science and technology have fabricated tiny semiconductor devices e.g. laser diode which produces a narrow beam of light in invisible bands (1280-1620 nm) or visible spectrum (400-700 nm). Optical wireless communication OWC is one of the most important applications of communication based on laser. Free space optics (FSO) is the only wireless technology having data rates in Gbps but vulnerable to weather conditions. FSO can be implemented in deep space due to unavailability of air in that atmosphere. Unlike Earth atmospheric conditions, there is no such attenuation factor present in space.

[1] Discuss the requirements to establish the optical inter-satellite link ISL for satellite constellations. [2] suggest many schemes to establish the inter-satellite link because high accuracy of laser beam positioning is essential to process laser ISL, it also discuss the different methods and architectures of free space optical communications. World first truly deep space lasercom (laser communication) was demonstrated by NASA and MIT Lincoln Laboratory in 2003. Their mission is to demonstrate optical wireless link of 3-50 Mbps between Mars telecom orbiter and NASA satellite. This laser communication duplex link has a round trip time of 40 minutes. In this mission space terminal, telescope receive array and ground terminal is developed by MIT Lincoln Laboratory. In deep space optical wireless communication systems can be used to communicate with the space shuttles and the satellites to get the better performance than the RF systems in terms of low cost, less power consumption of optical transmitter, low weight and with the small antenna aperture size. LLCD (Lunar Laser Communications Demonstration) is the project started by the NASA and MIT Lincoln Laboratory in 2009. The main goal of LLCD program is to demonstrate high performance optical wireless duplex link of 622 Mbps optical Downlink and 20 Mbps optical Uplink from a small antenna terminal [3-5].

Researchers and Engineers use amplified optical fiber transmitters, secure and power efficient encoding and decoding methods, optical MZ modulators, coherent receivers, FEC( forward error correction) techniques and optical preamplifier at the receiver end in order to extend the link range of lasercom [6]. Today the highest capacity communication satellite ViaSat-1has a data rate of 134 Gbps. ViaSat-1 is launched from Kazakhstan and it is in GEO orbit above North America [7]. At present 6967 satellites are in the orbit and out of 6967 satellites only 1328 active satellites are orbiting the earth [8]. Because of the communication requirements of observation satellites the numbers are increasing exponentially year by year.

[9] Model the IsOWC link at different wavelengths, data rates and ranges between satellites at LEO and concluded that performance of the IsOWC link is badly affected by increasing the data rate and distance between the links. [10] investigated and modeled the IsOWC systems between to LEO satellites at the data rate of 2.5 Gbps at the of 1000 Km. they compare 850 nm wavelength with 1550 nm wavelength and concluded that 850 nm wavelength gave better results in terms of SNR, Q factor and Total Power at low transmitting power.

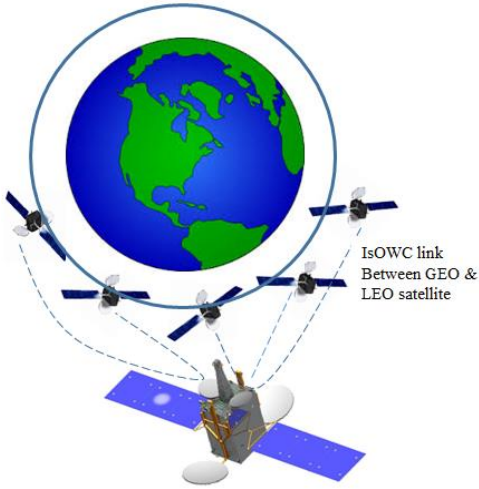

Fig. 1: model of inter-satellite optical wireless communiation link between GEO & LEO satellites

[11] modeled the IsOWC systems and investigated the system performance with and without square root module at the receiver end and concluded that acceptable BER and improved SNR can be achieved by using Square root module. This paper is presented the model simulation and performance analysis of inter-satellite optical wireless communication IsOWC system between GEO and LEO orbit satellites at the data rate of 10 Gbps over a space range of 40,000 Km. earlier research work did not exceed the data rate above 1 Gbps for GEO & LEO inter-satellite link (ISL). This research article is divided into three sections. Section 2 covers the system description and parametric configuration of inter-satellite link. Section 3 covers the results of IsOWC system and section 4 concludes the research work.

## II. INTER-SATELLITE MODEL DESCRIPTION

The IsOWC system is modeled in this paper which comprises of three basic communication components which are optical transmitter, propagation model and the optical receiver which is presented in figure 2. Inter-satellite link ISL is established between GEO and LEO satellite which is shown in figure 1. IsOWC system is used to communicate with each other in full-duplex mode and this system is designed and simulate in OPTI-SYSTEM simulator. This full-duplex system consists of two simplex systems; in order to investigate the IsOWC system performance single simplex system is studied. Where optical transmitter is in GEO satellite and optical receiver is in LEO satellite and optical light at infrared wavelength is used to communicate between inter-satellite links ISL. IsOWC is a kind of free space optics in which propagation medium is assumed as vacuum and all the attenuation factors due to atmosphere are considered as zero. OWC is a little dissimilar from optical fiber communication OFC in terms of propagation channel. Free space optics FSO is also abbreviated as lasercom (laser communication). Telemetry, tracking and control TT&C system offer fundamental communication path to and from the satellite and it is the only way to examine and control the satellite functions. TT&C subsystem of a satellite offer a link between the facilities on the Earth and the satellite itself and the main goal of TT&C function is to guarantee that the satellite is working properly. Irrespective of the application of the TT&C subsystem is essential for all satellites because it is a fundamental part of the spacecraft bus. Telemetry performs the operation of Health monitoring of the satellite and it collect, process and transmit the data from many satellites subsystems. Tracking and ranging is responsible for the satellite's precise location by transmitting, processing and receiving the ranging signals. Many of the satellite functions are automated and do not need ground station involvement. Proper control of a spacecraft via receiving the commands, processing and then implementation of the commands from the satellite or from the ground station are the major operations which are done by the control TT&C system. Transmitter of the optical signal receives the electrical data sequence from the TT&C system of satellite. NRZ pulse generator is used along with MZ-modulator which gets the input from the CW laser. EDFA optical amplifier is used to amplify the optical signal before the IsOWC link and then transmit the signal in space. In our proposed model we suppose that there is no meteorite or any other space dust particles in the path of optical signal. CW laser of power of 15 dBm and linewidth of 10 MHz is used in our suggested optical system. Optical antennas of transmitter and receiver has an aperture diameter of 20 cm and the gains of optical transmitter and receiver are zero. Optics efficiency of optical transmitter and receiver is equal to one. Optical transmitter and receiver antennas are supposed to be ideal and pointing error of both antennas is 0 urad. Propagation delay and other additional losses due to mispointing are also supposed as zero. Simulation Parameters of the IsOWC system is givern below in table 1.

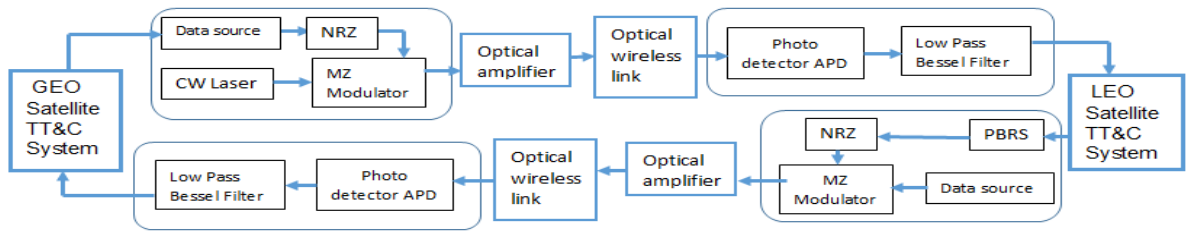

Fig. 1: Block diagram of IsOWC System: PBRS-Pseudo bit random sequence generator, NRZ-non return to zero, CW-Continuous wave laser, MZM-Mach zander modulator, TT&C-telemetry tracking and communication, APD-Avalanche Photo-detector.

TABLE 1: SIMULATION PARAMETERS OF ISOWC SYSTEM

| Frequency | 860 nm |
|---|---|
| Range | 40,000 Km |
| Data rate | 10 Gbps |
| Sequence length | 32 bits |
| Samples per bit | 64 |
| Number of Samples | 2048 |
| Extension Ratio of MZM | 30 dB |
| Dark current | 10 nA |

### III. RESULTS AND DISCUSSION

In reference simulation, wavelength of 860 nm is used in CW laser with the input power of 15 dBm. Line width of 10 MHz is used in Optical transmitter of reference simulation. Data sequence from the TT&C subsystem is passes through the NRZ pulse generator. MZ-modulator is used to externally modulate the optical signal. Optical amplifier is used before the OWC channel to amplify the optical transmitting signal with the gain of 30 dB and the noise figure of 4 dB. OWC is considered as attenuation free channel because of vacuum. Range of the OWC channel is of 40,000 Km with the antenna diameter of optical transmitter and receiver terminal is of 20 cm. On the optical receiver side, Avalanche photo detector is used to detect the light with the ionization ratio of 0.9. Low pass Bessel filter with the cut-off frequency of 0.75*bit rate with the order 4 is used to filter the electrical signal. After filtering, the electrical signal is fed into the TT&C subsystem of LEO satellite to analyze the signal quality, signal power, BER and jitter in the signal. Figure 3 shows the eye diagram of the reference simulation, it has the Q factor of 30 and the BER of 2.933e-201. Eye diagram has the height of 7.133e-005.

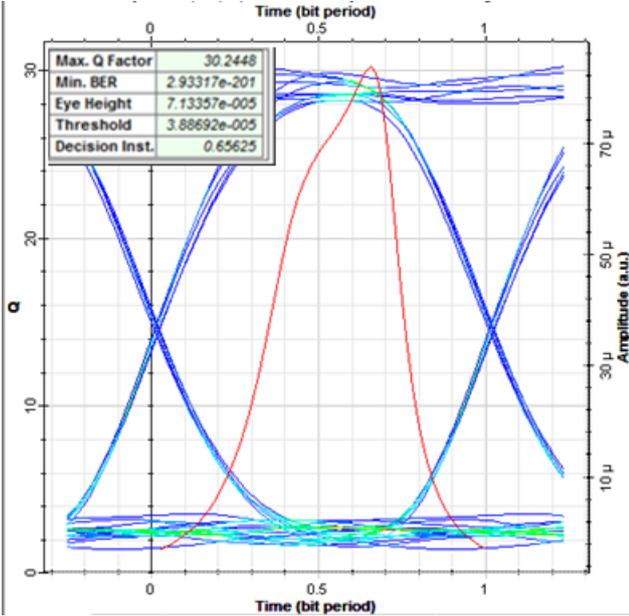

Fig. 3: Eye diagram of reference simulation of IsOWC system at the wavelength of 860 nm.

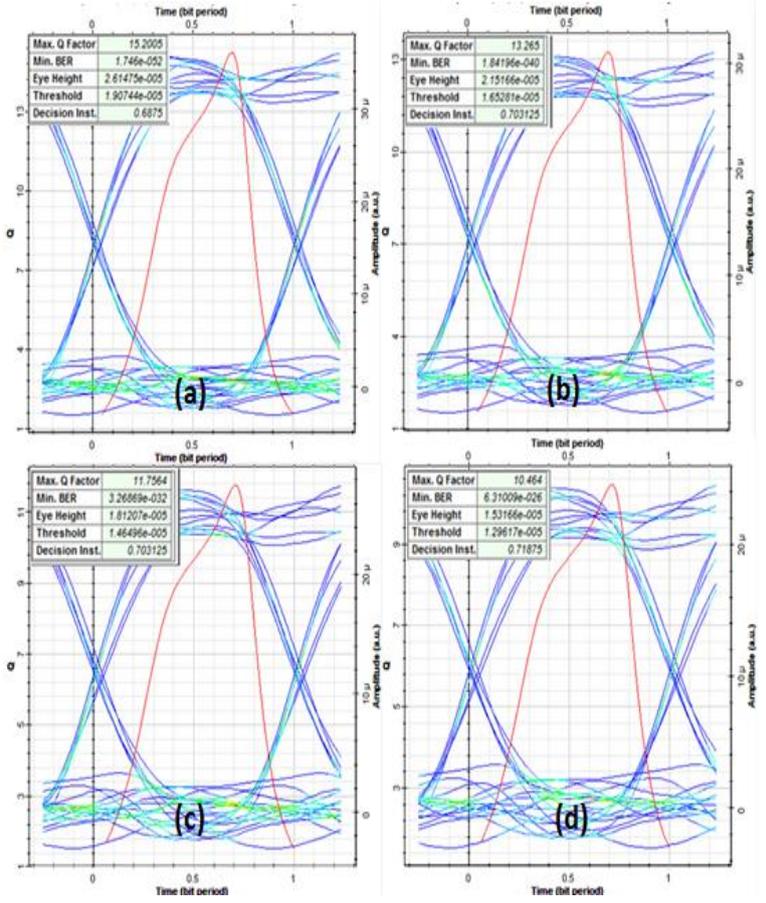

Fig. 4: (a) shows the eye diagram of IsOWC system at wavelength of 1340 nm, (b) shows the eye diagram of IsOWC system at wavelength of 1450 nm, (c) shows the eye diagram of IsOWC system at wavelength of 1550 nm, (d) shows the eye diagram of IsOWC system at wavelength of 1650 nm.

By keeping all the parameters constant, we change the wavelength of reference simulation from 860 nm to 1450 nm, 1340 nm, 1550 nm and 1650 nm. It is seen from figure 4 that with the increase in optical signal wavelength the signal quality factor and total signal power decreases. BER and jitter of the IsOWC system increases gradually as shown in figure 4. At wavelength of 1340 nm, the Q-factor of the system decreased to 14.2005 and BER of the IsOWC increased to 1.746e-52. Where signal power is also decreased to -64.9 dBm as shown in figure 4(a). At wavelength of 1450 nm, the Q-factor of the system decreased to 13.265 and BER of the IsOWC increased to 1.84e-040. Where signal power is also decreased to -66.27 dBm as shown in figure 4(b). At wavelength of 1550 nm, the Q-factor of the system decreased to 11.75 and BER of the IsOWC increased to 3.26e-032. Where signal power is also decreased to 67.43 dBm as shown in figure 4(c). At wavelength of 1650 nm, the Q-factor of the system decreased to 10.464 and BER of the IsOWC increased to 6.31e-026. Where signal power is also decreased to -68.5201 dBm as shown in figure 4(d).

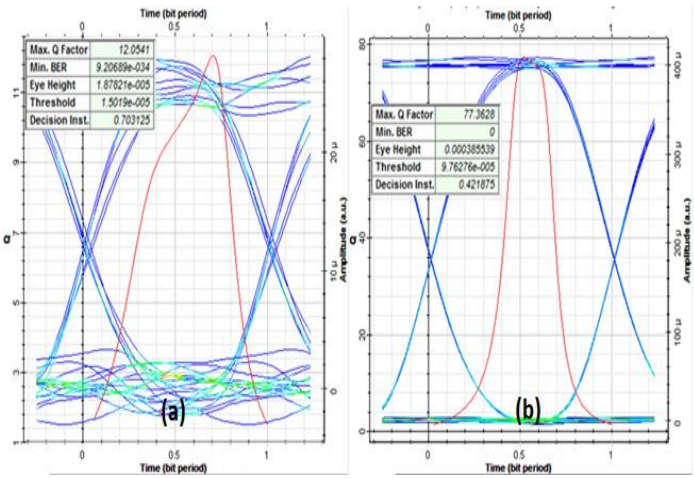

Fig. 5: (a) shows the eye diagram of IsOWC system with the aperture diameter of antenna is 15 cm, (b) shows the eye diagram of IsOWC system with the aperture diameter of antenna is 30 cm.

Antenna aperture diameter of transmitter and receiver of both satellites has the significant impact on the satellite speed and drag. Because by increasing the aperture diameter of antenna the mass, drag and size of the satellite payload increases. In this paper, except antenna aperture diameter all the parameters are kept constant. By decreasing the antenna aperture diameter from 20 cm to 15 cm, as a result the mass, size and drag of the satellite decreases which is a good sign but by decreasing the aperture diameter the Q-factor and the signal power is also decreased to 12.05 and -67.19 dBm as shown in figure 5 (a). Figure 5 (b) shows the eye diagram of the IsOWC system simulation in which antenna aperture diameter is 30 cm. by increasing the aperture diameter although the Q-factor and the signal power is increased as shown in figure 5 (b) but size, mass and drag of the payload is also increased.

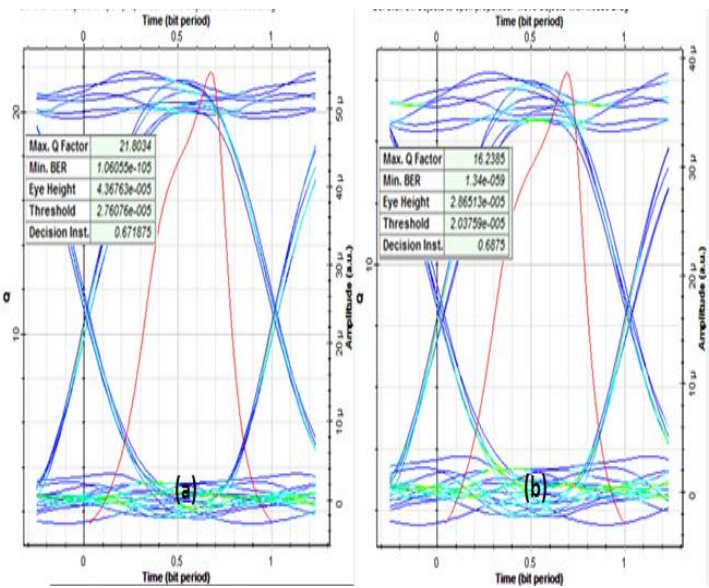

Fig. 6: (a) shows the eye diagram of the IsOWC system at the distance of 50,000 Km. (b) shows the eye diagram of the IsOWC system at the distance of 60,000 Km.

By keeping all the parameters, range of the ISL is changed to see the effect of distance on the IsOWC system. To see the effect of distance, system is simulated at 50,000 km and at 60,000 Km. it is noticed that by increasing the distance of ISL the quality and total signal power of the optical signal is decreased. By increasing the distance from 40,000 Km to 50,000 Km the signal power is decreased to -61.08 dBm and the Q-factor is decreased to 21.80 as shown in figure 6 (a). When the distance is further increase to 60,000 Km the Q-factor is deceased 16.23 and the total signal power is decreased to -64.24 dBm.

CONCLUSION

In this research article, IsOWC system is established between LEO and GEO satellite over the distance of 40,000 Km and at the data rate of 10 Gbps. From this simulation it is concluded that system has the better Q-factor and the total signal power in case of 860 nm rather than 1450 nm, 1550 nm or 1650 nm. It is also concluded that aperture diameter of the telescope and the range of the ISL link has a significant effect on the quality of the optical signal.